# Investigation of dielectric anomalies at cryogenic temperatures in the (1-x)[Pb(Mg$_{1/3}$ Nb$_{2/3}$)O$_3$]-xPbTiO$_3$ system

M. H. Lente,[1] A. L. Zanin, E. R. M. Andreeta, I. A. Santos, D. Garcia and J. A. Eiras

Universidade Federal de São Carlos - Departamento de Física - Grupo de Cerâmicas Ferroelétricas CEP 13565-670 - São Carlos - SP – Brazil

Complex electrical permittivity measurements in (1-x)[Pb(Mg$_{1/3}$ Nb$_{2/3}$)O$_3$]-xPbTiO$_3$ ceramics for $0.10 \leq x \leq 0.40$ were performed in the frequency and temperature range from 1 kHz to 100 kHz and from 15 to 600 K, respectively. Unexpected dielectric anomalies, which were frequency dependent, were found at cryogenic temperatures. It was verified that this dielectric behavior is always observed, independently of whether the composition presents a "normal" ferroelectric or a relaxor characteristic. The results were analyzed within the framework of two current models found in the literature. It was concluded that none of the proposed mechanisms uniquely satisfies the experimental results in the composition range investigated here.

[1] E-mail address: mlente@df.ufscar.br

Research on the nature of dielectric response of ferroelectric materials has been doubtless a long-standing puzzle, and has motivated intensive theoretical and experimental studies. Indeed, the investigation of the temperature dependence of the electrical permittivity in ferroelectric systems has been revealed as a helpful tool to study several physical phenomena such as structural phase transitions [1, 2, 3] and dielectric relaxations.[4, 5, 6] For instance, the nature of the dielectric response in relaxor ferroelectrics (relaxors) has been one of the most challenging problems in the physics of ferroelectricity. Relaxors are characterized by (i) frequency dependence of their permittivity peak (frequency relaxation) with freezing of the relaxation time spectrum, (ii) nonergodic behavior at low temperatures and (iii) long range order induced by a dc electric field.[7]

In addition, unexpected frequency dependent dielectric anomalies at cryogenic temperatures have also been reported in different relaxor systems.[4, 5, 6] Two models have been proposed in the current literature to explain the physical origin of this low temperature dielectric behavior. The first one assumes that chemical inhomogeneity destroys local translational symmetry leading to thermally agitated local-polarization fluctuations, which would be responsible for the relaxations.[4] The second one suggests that the anomalies in the dielectric loss observed at low temperatures are dominated by structural irregularities (fractal clusters) inside normal micron-sized domains.[5, 8] Although several works have been dedicated to investigating the dielectric properties of ferroelectric systems in order to obtain a better understanding of their MPB, the dielectric properties at very low temperatures have not received much attention.

The objective of this work is to investigate the electrical permittivity of $(1-x)[Pb(Mg_{1/3}Nb_{2/3})O_3]-xPbTiO_3$ ceramics for $0.10 \leq x \leq 0.40$ at temperatures down to 15 K. The results reveal remarkable dielectric anomalies with frequency dispersion at cryogenic temperatures over all composition ranges investigated. The physical origin of these anomalies will be discussed in the framework of the models proposed in the literature.

(1-x)[Pb(Mg$_{1/3}$Nb$_{2/3}$)O$_3$]-xPbTiO$_3$ (PMN-PT) ceramics with x = 0.10, 0.35 and 0.40 were prepared through the columbite method and densified by a uniaxial hot-pressing technique. To assist the reader, the phase diagram of PMN-PT is shown in Fig. 1.[9] Synthesis and processing details can be found elsewhere.[10,11] X-ray diffraction analysis of the crushed ceramic powders revealed only perovskite phase for all compositions. The relative apparent density was higher than 98% of the theoretical density. The sintered ceramic bodies were cut into a bar shape of 5 x 4 mm$^2$ and polished to a thickness of 0.5 mm for dielectric measurements. After that, the samples were annealed at 900 K for 1 h and gold electrodes were sputtered onto the sample surfaces. Dielectric measurements (1 kHz -100 kHz) were made through an automated system using an Impedance Analyzer HP4194A. The temperature range investigated was from 15 K to 600 K using a cryogenic refrigeration system (APD Cryogenics Inc.) and a home-made furnace at a constant cooling rate of 2 K/min.

Figures 2 (a)-(f) show the real ($\varepsilon'$) and imaginary ($\varepsilon''$) components of the relative electrical permittivity as function of the temperature and frequency for all compositions, which are located in the rhombohedral (x=0.10), monoclinic (x=0.35) and tetragonal (x=0.40) regions of the PMN-PT phase diagram (see Fig. 1). The column on the left shows the measurements performed in unpoled samples, while the column on the right shows the results for respective poled samples. The samples with x = 0.40 and 0.35 were poled by applying a field of 25 kV/cm at 350 K for 30 min, whereas for x = 0.10 the measurement was performed under a bias electric field of 5 kV/cm. The insets show the measurements at high temperatures. The electrical permittivity of the unpoled samples shows a maximum at a temperature T$_m$ that is in good accordance with that predicted in the phase diagram for the respective paraelectric- ferroelectric phase transition (Fig. 1). A dielectric anomaly associated with the tetragonal-monoclinic phase transition for composition x = 0.35 was seen only after the poling. This anomaly is also in accordance with the PMN-PT phase diagram. However, the most remarkable result revealed in the dielectric measurements is that all compositions presented clear frequency dependent dielectric anomalies (or relaxations) at temperatures lower than 150 K. Moreover, according to the phase diagram of PMN-PT (Fig. 1) these anomalies cannot be associated with a phase transition. In general, these anomalies are better visualized in the dielectric loss and in the poled or electric biased samples.

As commented on previously, similar anomalies have been found in some relaxor systems and two models have been proposed in the literature to explain them. The assumption of polarization fluctuation would imply that the application of a dc electric field (field cooling or similar electrical boundary condition such as poling of the sample) could freeze-in local-polarization fluctuations, thus suppressing dielectric relaxations. [4] However, our results show clearly that in both cases the dielectric relaxation still persists and becomes more evident. Moreover, the assumption of the polarization fluctuation model implies the supposition of a break of local symmetry that leads to the relaxor character. [4,12] Nevertheless, our results reveal that the low temperature dielectric relaxation is noticed even in the composition with a "normal" ferroelectric characteristic [x=0.40, see insert in Fig. 2(e)], where in principle there is no such internal heterogeneity.

Some works have also pointed out that structural irregularities (or clusters) are the main cause of the above-mentioned dielectric anomalies. [5,8] In this situation, it is supposed that a ferroelectric system can be frozen into a state with clusters of the low temperature state embedded in the average symmetry of the high temperature state. [8] Nevertheless, these investigations were conducted in ferroelectric materials with compositions in the vicinity of the Morphotropic Phase Boundary (MPB), where normal micron-sized domains can coexist with structural irregularities.[8] The phase coexistence is also confirmed in PMN-xPT for 0.31

≤ x ≤ 0.37 at temperatures down to 20 K.[9] Therefore, this coexistence could justify the dielectric anomalies found for the composition with x = 0.35, assuming the structural irregularities as the origin of the dielectric anomalies. Nevertheless, no phase coexistence is found for x ≤ 0.30 and x ≥ 0.39 [9] but dielectric relaxation is still noticed in this composition range, as shown in Figs. 2. Particularly for x = 0.10, were found only nanopolar regions not micron-sized domains.[13] Therefore, the supposition that the dielectric anomalies at cryogenic temperatures are due to structural irregularities coexisting with micron-sized domains cannot be extended to the PMN-PT for x ≤ 0.30 and x ≥ 0.39.

At first glance, it seems to be very difficult to find a unique mechanism able to explain the dielectric relaxations at cryogenic temperatures in the composition range investigated here (0.10 ≤ x ≤ 0.40). This difficulty arises since both relaxor (short-range order) and normal ferroelectric (long-range order) characteristics are achieved in this composition interval and, therefore, the supposed mechanism should be valid in both distinct situations. Consequently, the above discussed models do not seem able in principle to explain the dielectric results in the composition range investigated here, although they can be valid in a specific range, as originally proposed. Undoubtedly, these results reveal interesting features and demand further investigation in order to determine the mechanism responsible for such dielectric relaxations in the whole extension of the PMN-PT phase diagram.

In summary, it was reported in this letter the observation of dielectric anomalies with relaxor-type behavior at cryogenic temperatures in PMN-xPT ceramics for 0.10 ≤ x ≤ 0.40. The results revealed that the dielectric relaxation is always observed, independently of wheter the composition presents a "normal" or a relaxor characteristic. It was not possible to find a unique mechanism that explains the results for the whole extension of the PMN-PT phase diagram.

The authors thank FAPESP, CAPES and CNPq for financial support.

# Figure Captions

Figure 1: Phase diagram of $(1-x)[Pb(Mg_{1/3}Nb_{2/3})O_3]$-$xPbTiO_3$.

Figure 2: Temperature dependence of the real ($\varepsilon'$) and imaginary ($\varepsilon''$) components of the relative electrical permittivity of PMN-xPT for x = 0.10, 0.35 and 0.40. The column on the left shows the measurements performed in unpoled samples, while the column on the right shows the results for the respective poled samples. For x = 0.10 in Fig. 2(b) the measurement was performed under a bias field of 5 kV/cm.

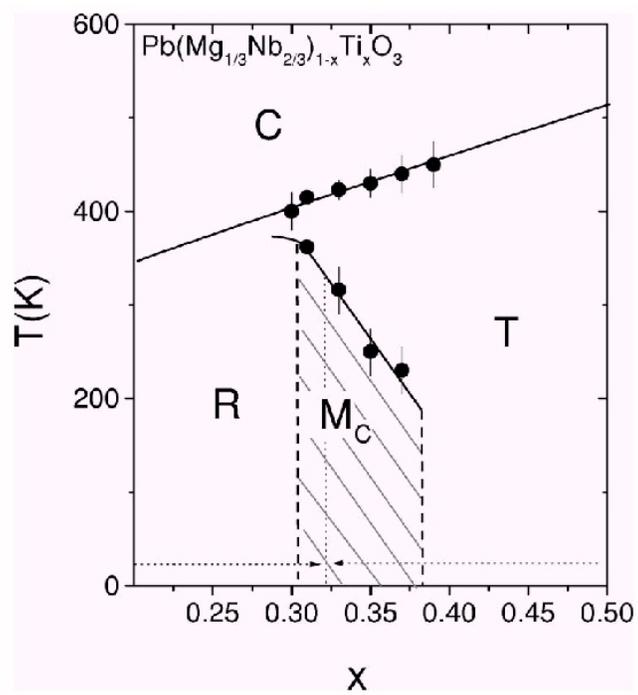

Figure 1: M. H. Lente et al.

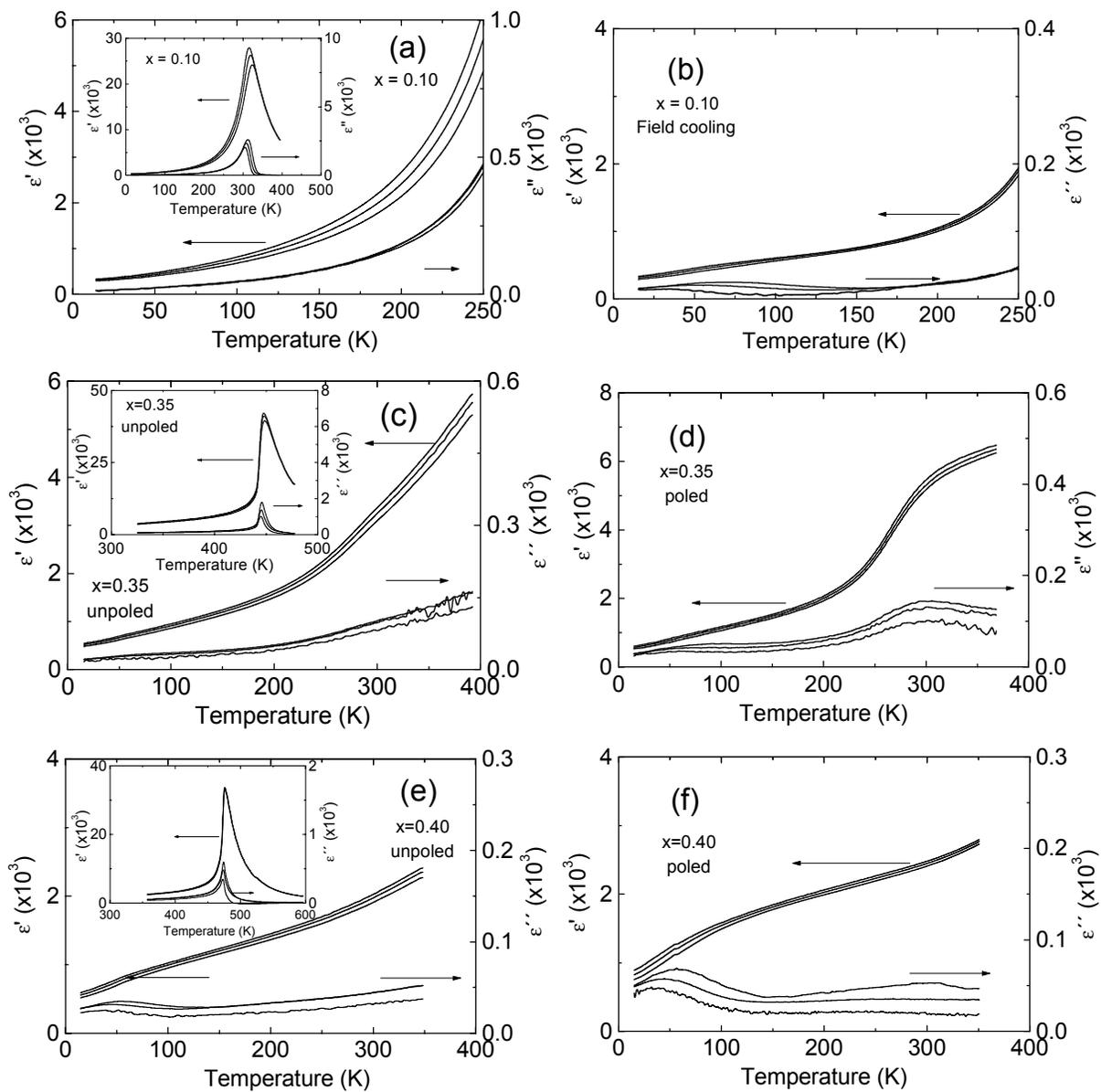

Figure 2: M. H. Lente